 \documentclass[final,5p,times,twocolumn]{elsarticle}

\usepackage{lineno,hyperref}

\usepackage{color}
\usepackage{mwe}
\usepackage{floatrow}
\usepackage{amssymb}
\usepackage{amsmath}
\usepackage{soul}
\usepackage[normalem]{ulem}

\DeclareMathOperator{\erfc}{erfc}

\newcommand{\inches}{\ensuremath{{}^{\prime\prime}}}

\def\be{\begin{equation}}
\def\ee{\end{equation}}
\def\bi{\begin{itemize}[noitemsep,leftmargin=*]}
\def\ei{\end{itemize}}
\def\bmx{\begin{matrix}}
\def\emx{\end{matrix}}
\def\ptype{\emph{p}-type}
\def\ntype{\emph{n}-type}
\def\pside{\emph{p}-side}
\def\nside{\emph{n}-side}

\journal{Nuclear Instruments and Methods in Physics Research A}

\begin{document}

\begin{frontmatter}

\title{Fabrication of low-cost, large-area prototype Si(Li) detectors for the GAPS experiment}

\author[MIT]{Kerstin Perez\corref{mycorrespondingauthor}}
\cortext[mycorrespondingauthor]{Corresponding author}
\address[MIT]{Massachusetts Institute of Technology, Cambridge, MA 02130, USA}

\author[SLAC]{Tsuguo Aramaki}
\address[SLAC]{Stanford Linear Accelerator Center, Menlo Park, CA 94025, USA}

\author[CU]{Charles J. Hailey}
\address[CU]{Columbia University, New York, NY 10027, USA}

\author[CU]{Rachel Carr}
\author[MIT]{Tyler Erjavec}
\author[JAXA]{Hideyuki Fuke}
\address[JAXA]{Institute of Space and Astronautical Science, Japan Aerospace Exploration Agency (ISAS/JAXA), Sagamihara, Kanagawa 252-5210, Japan}
\author[CU]{Amani Garvin}
\author[CU]{Cassia Harper}
\author[CU]{Glenn Kewley}
\author[CU]{Norman Madden}
\author[CU]{Sarah Mechbal}
\author[MIT]{Field Rogers}
\author[CU]{Nathan Saffold}
\author[CU]{Gordon Tajiri}

\author[Shimadzu] {Katsuhiko Tokuda}
\address[Shimadzu]{Shimadzu Corporation, Kyoto, Kyoto 604-8511, Japan}

\author[CU]{Jason Williams}
\author[Shimadzu]{Minoru Yamada}

\begin{abstract}
A Si(Li) detector fabrication procedure has been developed with the aim of satisfying the unique requirements of the GAPS (General Antiparticle Spectrometer) experiment. 
Si(Li) detectors are particularly well-suited to the GAPS detection scheme, in which several planes of detectors act as the target to slow and capture an incoming antiparticle into an exotic atom, as well as the spectrometer and tracker to measure the resulting decay X-rays and annihilation products.
These detectors must provide the absorption depth, energy resolution, tracking efficiency, and active area necessary for this technique, all within the significant temperature, power, and cost constraints of an Antarctic long-duration balloon flight.
We report here on the fabrication and performance of prototype 2\inches-diameter, 1-1.25\,mm-thick, single-strip Si(Li) detectors that provide the necessary X-ray energy resolution of $\sim4$\,keV for a cost per unit area that is far below that of previously-acquired commercial detectors.
This fabrication procedure is currently being optimized for the 4\inches-diameter, 2.5\,mm-thick, multi-strip geometry that will be used for the GAPS flight detectors.
\end{abstract}


\end{frontmatter}


\section{Introduction}

Lithium-drifted silicon (Si(Li)) detectors were first developed in the 1950s and 1960s~\cite{Pell:1959,Goulding:1966}. 
To fabricate these detectors, Li ions are ``drifted" through \ptype\ Si in order to produce well-compensated, almost intrinsic regions with thicknesses of up to many millimeters. 
Si(Li) detectors have previously been investigated for use in Compton telescopes~\cite{Hau:2003,Kurfess:2007,Harkness:2013}, X-ray satellite telescopes~\cite{Serlemitsos:1990}, ionizing particle detection~\cite{POPEKO2008235}, neutrino-electron scattering experiments~\cite{Popeko:2004}, and low-energy X-ray measurements~\cite{JAKLEVIC1988598}. 

Despite these successes, recent improvements in Si processing and evolving experimental needs warrant revisiting Si(Li) fabrication techniques. 
Early studies in particular used poorer quality Si than is now widely available, 
and as a result were restricted to fabricating detectors with small ($\sim$1\,cm diameter) active areas~\cite[e.g.,][]{Murray1965,Baron1966}. 
Achieving good energy resolution in larger-area Si(Li) detectors is challenging due to the increased leakage current and capacitance. 
As a result, detectors with active areas more than several centimeters in diameter were initially restricted to measurements of $\alpha$ particles, protons, or pions \cite[e.g.,][]{Nakamoto1975,Yoshimori1975,Miyachi1987}, since the large energy deposited by these heavy charged particles compensates for these effects. 
As large-area, high-resistivity Si became available, detectors with active areas several centimeters in diameter achieved $\sim1$\,keV energy resolution for X-rays~\cite[e.g.,][]{Pehl1986,Fong1982}. 
These detectors, however, required operation at almost liquid nitrogen temperature.
	
We report here on our work adapting Si(Li) detector fabrication techniques to satisfy the 
large-area, high-yield, low-cost, and relatively high-temperature (-35 to -45\,C) requirements of the GAPS (General Antiparticle Spectrometer) Antarctic balloon experiment~\cite{Hailey:2009,Hailey2013290,Aramaki2015}. 
The Antarctic GAPS program will be the first experiment optimized specifically for detection of low-energy ($E < 0.25$\,GeV/n) cosmic antideuterons and antiprotons as signatures of dark matter annihilation or decay. 
A first-time detection of cosmic antideuterons would be an unambiguous signal of new physics, providing a new avenue to probe a variety of dark matter models that is complementary to collider, direct, or other cosmic-ray searches, and would open a new field of cosmic-ray research.
GAPS will also provide a precision antiproton spectrum in a low-energy region currently inaccessible to any experiment, providing sensitivity to light dark matter as well as cosmic ray propagation models, which are crucial to interpreting any charged cosmic ray signal. 

To achieve sensitivity to cosmic antinuclei in this unexplored low-energy range, GAPS will use Si(Li) detectors for a novel particle identification technique based on exotic atom capture and decay. 
First, the low-energy incident cosmic-ray antiparticle is trapped by the target Si and forms an exotic atom. 
The signature of characteristic de-excitation X-ray energies and nuclear annihilation decay product multiplicity, along with the incident $dE/dx$ losses and stopping depth within the instrument, is then used to distinguish different antiparticle species. 

The success of this detection technique for discovery of a rare cosmic antideuteron signal relies on both large active detector area and high background rejection capabilities, within the temperature, power, and cost constraints of an Antarctic long-duration balloon flight program.
The target antideuteron flux sensitivity of $2\times10^{-6}$\,m$^{-2}$\,s$^{-1}$\,sr$^{-1}$\,(GeV/n)$^{-1}$ (99\% C.L.) requires $\sim$10\,m$^2$ of detector area distributed over ten planes~\cite{Aramaki2015}. 
Development of a cost-effective detector fabrication process utilizing an affordable Si substrate is key to enabling this large sensitive area.
The 2.5\,mm thickness of these detectors is optimized to provide high escape fractions for the exotic atom X-rays, in order to allow for detection in the surrounding Si, but also to provide sufficient total instrument depth to stop incoming antinuclei with energies up to 0.25\,GeV/n.
Each detector will be segmented into readout strips of width $\sim$1--3\,cm, in order to distinguish tracks from incident particles, exotic atom X-rays, and nuclear annihilation products.
As the X-ray energies are uniquely determined by the antiparticle species, each strip must deliver $\sim$4\,keV X-ray energy resolution in order to further distinguish antideuterons from antiprotons.
The energy resolution performance is complicated by the temperatures achievable within the power and weight constraints of an Antarctic balloon flight, which impose a relatively high operating temperature of -35 to -45\,C.
In addition, the energy resolution scales with the detector capacitance, but due to the power constraints of a balloon flight, the total number of readout channels is limited and thus we cannot reduce the detector capacitance by having more, smaller, readout strips. 
In recent years, high-purity Si has become available at a low enough cost to make high-volume detector fabrication feasible without lithium drifting; however, the $\sim$2.5\,mm depletion depth required for GAPS would require an applied bias of several thousand volts, and weight constraints preclude the use of a pressure vessel for the detector volume. 
The GAPS flight Si(Li) detectors can operate at an applied bias of 250 V and will thus require less power and be significantly less prone to high voltage breakdown at the very low atmospheric pressure of a balloon flight.

The prototype GAPS (pGAPS~\cite{Doetinchem2013pGAPS,2014NIMPA.735...24M,2014AdSpR..53.1432F}) flight, launched from Taiki, Japan in 2012, included six 4\inches-diameter, 2.5\,mm-thick Si(Li) detectors acquired from SEMIKON Detector GmbH (now defunct)~\cite{2012NIMPA.682...90A}.
The circular active region of these detectors was segmented into 8 strips of equal area via straight ion-milled grooves.
This was the first experiment to validate the use of Si(Li) detectors in an ambient balloon environment.
To date, these are the only large-area Si(Li) detectors shown to deliver energy resolution of $\sim$3-4\,keV at temperatures as high as -35\,C~\cite{Aramaki2013}.
As SEMIKON used a proprietary commercial procedure, and as the company is no longer in business, appropriate fabrication methods must be re-developed. 
In addition, SEMIKON used Si substrates from only Topsil Semiconductor Materials. 
In partnership with SUMCO Corporation, an alternate Si substrate has been developed.  
The decrease in cost of the SUMCO wafers compared to Topsil and the successful validation of this material in the prototype detectors are key to enabling the GAPS mission.

\begin{figure}[tp]
\vspace{-0.in}
\includegraphics[width=0.95\linewidth]{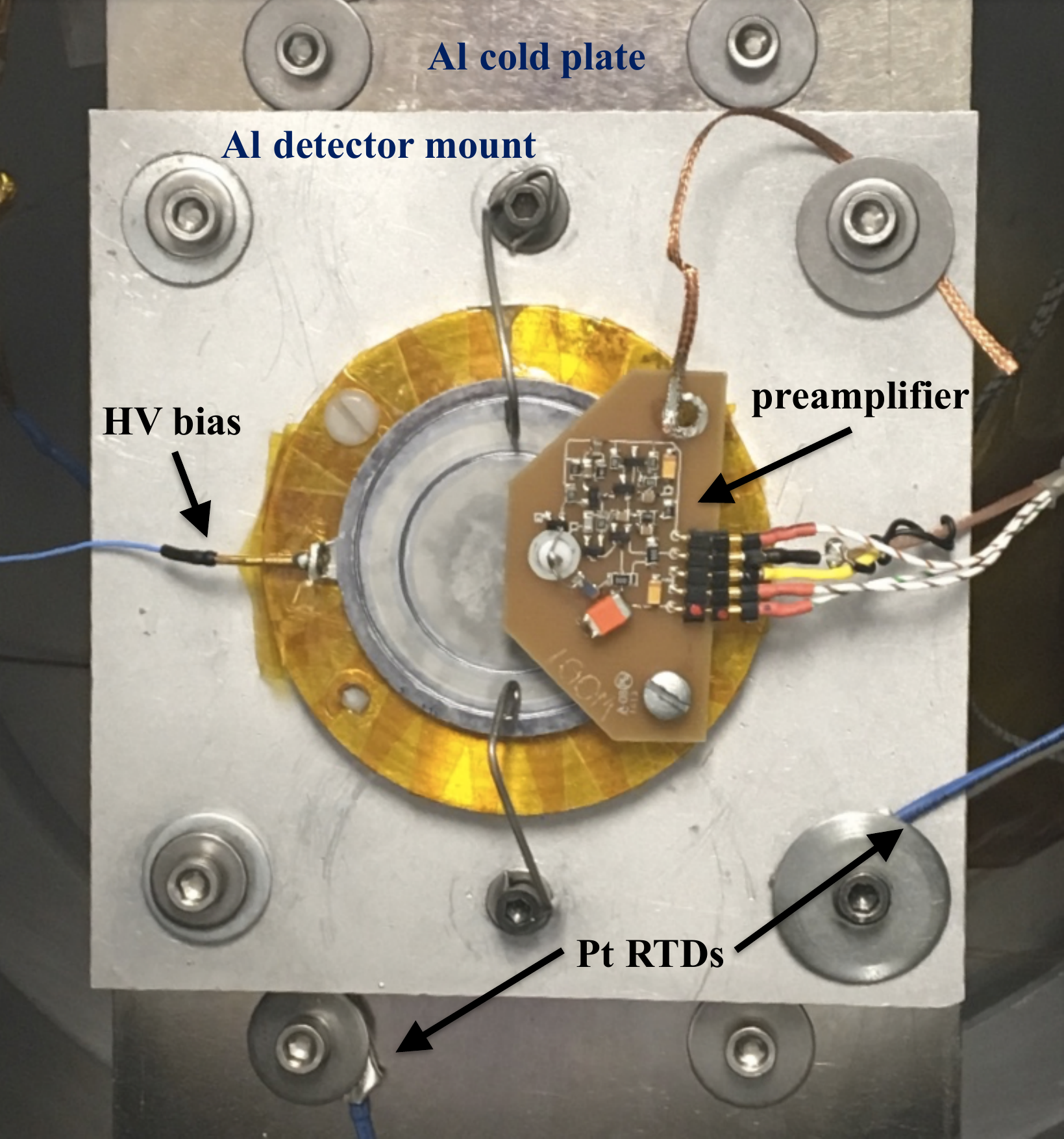}
\caption{\label{fig:testphoto} Prototype 2\inches-diameter, 1.25\,mm-thick detector with 1.25\inches-diameter active area, mounted with a custom charge-sensitive preamplifier.}
\vspace{-0.in}
\end{figure}

In this paper, we report the development of prototype detectors made from 2\inches-diameter Si wafers, with final active areas of 1.25\inches-diameter and $\sim$1-1.75\,mm thickness, that satisfy the GAPS performance requirements.
An example of such a detector is shown in Fig.~\ref{fig:testphoto}. 
In partnership with Shimadzu Corp., similar methods are now being finalized to fabricate the 4\inches-diameter ($\sim$3.5\inches-diameter active region), 2.5\,mm-thick, multi-strip detectors that will be used for the initial Antarctic GAPS flight, with production runs beginning in 2018.
The performance of these first flight-geometry Si(Li) detectors will be detailed in a future publication.

The paper proceeds as follows.
In Sec.~\ref{sec:silimethod}, we provide an overview of the essential features of Si(Li) detector fabrication.
Sec.~\ref{sec:prototype} describes the details of the method developed for the GAPS prototype detectors, and 
Sec.~\ref{sec:performance} summarizes the prototype detector performance achieved by this method. 
Conclusions and directions of future work are presented in Sec.~\ref{sec:future}. 

\section{Overview of Si(Li) fabrication method}
\label{sec:silimethod}

Our procedure is based on the lithium-drifted guard ring detectors that were first developed in the 1960s~\cite{Pell:1959,Goulding:1966}. 
This method relies on two key properties of lithium: it is an easily-ionized donor atom in Si (0.033\,eV) and its ions have high mobility in the silicon lattice. 
Since the 1960s, this framework has been repeatedly modified and updated to suit a variety of applications. 
The procedure we have developed is described in Sec.~\ref{sec:prototype}, where we detail which methods do, and do not, work for our specific goals.
We first summarize here the basic principles and comment on some aspects that have warranted further study.

Li is first applied to the front surface of \ptype\ (free positive charge dominated) Si and diffused through a short depth. 
The Li atoms donate electrons, resulting in an \emph{n}+ (free negative charge dominated) Si lattice layer and residual free positive Li ions.
The depth of the resulting \emph{n}+ material is given by the $x$, in cm, that satisfies:  
\begin{equation}
\label{eqn:diffusion}
N_A = N_0 \erfc[\frac{x}{2(D t)^{1/2}}].
\end{equation}
Here, $N_A$ is the \ptype\ acceptor impurity (typically boron) concentration, $N_0$ is the lithium concentration at the surface, and $t$ is the diffusion time in seconds. $D$ is the diffusion constant, given by: 
\begin{equation}
D = 0.0023 \cdot exp(\frac{-7700}{T}) 
\end{equation}
in units of cm$^2$\,s$^{-1}$, with $T$ the diffusion temperature in Kelvin~\cite{Pell:1959}. 
 
Under reverse bias, the positive Li ions move away from the \ntype\ region, compensating the acceptor atoms in the \ptype\ bulk, as well as any intrinsic impurities in the Si. 
Throughout this drift process, the electric field adjusts to oppose the buildup of charge in any one region. 
For modest drift temperatures and voltages, large intrinsic (carrier-free) detector regions can be produced. 
The drifted width, $W$, after time $t$ seconds at an applied bias of $V$ in Volts, is given by 
\begin{equation}
\label{eqn:drifting}
W = (2 V \mu_{Li} t)^{1/2}
\end{equation}
Here, $\mu_{Li}$ is the Li ion mobility at a temperature $T$ in Kelvin:
\begin{equation}
\mu_{Li} = \frac{26.6}{T} exp(\frac{-7500}{T})
\end{equation}
in units of cm$^2$\,V$^{-1}$\,s$^{-1}$.
For thick wafers, either a circular groove is machined some distance from the outer edge of the wafer (producing an ``inverted-T" geometry) or several millimeters of material are removed from the circumference of the \nside\ of the wafer (producing a ``top-hat" geometry)~\cite{1966ITNS...13...93L} before drifting.
Such geometries allow for the drifting bias to be applied only to the interior of the wafer face, establishing an electric field that ensures that Li drifts through the bulk, not along the side, of the detector. 

The drifting procedure critically affects the final detector performance, including both the charge collection efficiency and bulk leakage current.
A wide array of methods, with varying reliability, have been used to indicate the successful end of the drift.
In some methods, the drift process is terminated when the leakage current increases sharply, indicating that the intrinsic region is approaching the \pside ~\cite[e.g.,][]{Ristenen1967}.
As any excess of Li on the \pside\ causes very high leakage current, any over-drifted material is then removed via lapping and chemical etching.
This method has primarily been validated using small ($\sim1$\,cm-diameter, 1-10\,mm-thick) detectors~\cite[e.g.,][]{Goulding:1966,Murray1965,Walton1993}. 
Larger-area detectors have been fabricated by applying boron-implanted contacts to the \pside\ before drifting. 
In this case, there is  no need to remove over-drifted material~\cite[e.g.,][]{Walton1978,Protic:2005} and the intrinsic region width is confirmed using $\alpha$ particle energy measurements. 
Other methods include ending the drift after a time calculated from Eqn.~\ref{eqn:drifting}~\cite[e.g.,][]{Landis1989}, ending the drift early and then removing undrifted material~\cite[e.g.,][]{Baron1966}, or ending the drift when the measured capacitance corresponds to the desired drifted width~\cite[e.g.,][]{Keffous2005}. 
As each method has previously been utilized for a different detector application, any approach we use must first be validated to meet the GAPS performance requirements.

The Li drift can be followed by a cleanup drift at a temperature significantly lower than the original drift, but significantly higher than the detector operating temperature. 
This allows the Li ions to redistribute at a temperature at which the mobile carrier density is very low, thus reducing any localized charge over-densities~\cite{Lauber:1969}. 
This practice also reduces the over-density of Li, known as the Li ``tail", that diffuses below the concentrated \emph{n}+ region during drift~\cite{Protic:2003,Protic:2005}. 

The guard ring geometry is key to reducing noise in Si(Li) detectors with large active areas.
After drifting, a circular groove is machined into the \emph{n}+ side so that the active area, which is inside the groove, can be electrically isolated from the guard ring, which is outside the groove. 
During operation, the bias is applied across the active region, while the guard ring is grounded.  
At high biases, the electric field forms a depletion region along the groove between the guard ring and the active region, 
causing effective resistances of over 100\,M$\Omega$~\cite{Goulding:1966,1964ITNS...11..221L}. 
For a large-area detector, the surface leakage current along the perimeter of the wafer can be many orders of magnitude larger than the bulk leakage current. 
With the guard ring, only surface leakage current originating in the narrow groove can make it to the detector readout. 
By reducing this surface leakage current, the guard ring structure significantly reduces detector noise.

In order to effectively isolate the active area from the guard ring, the groove must be deeper than the \emph{n}+ Li diffused layer and any Li tail. 
A thick initial \emph{n}+ layer is necessary to provide sufficient Li ions to compensate the full detector volume;
however a thinner \emph{n}+ layer can be implemented after drifting and prior to cutting the guard ring and any readout segmentation. 
This thinner \emph{n}+ layer can be achieved by either removing the entire initial Li layer via lapping and then diffusing a shallower Li region~\cite{Walton1978}, or by successive thinning of the original Li layer and verifying that sufficient density remains by measuring the sheet resistivity~\cite{Protic:2005}. 
In previous work, this thinner \emph{n}+  layer has been primarily motivated by the need to implement fine position elements using grooves that are only tens of microns wide and similarly shallow.  
Although strip widths of $\sim$1--3\,cm are sufficient for GAPS particle tracking, we are currently investigating if, additionally, such narrow and shallow grooves reduce the surface leakage current originating in the guard ring groove. 

Finally, electrodes are applied to the \nside\ and \pside\ of the detector. 
On the \nside, an Ohmic contact is applied to the active area and guard ring in order to ensure good charge collection efficiency with low series resistance. 
In contrast, a Schottky barrier contact is necessary on the \pside\ in order to prevent charge injection from the metal contact directly into the intrinsic detector region.

\section{GAPS prototype Si(Li) detector fabrication}
\label{sec:prototype}

We describe below the basic fabrication procedure developed for the GAPS prototype Si(Li) detectors, which serves as the basis of the method under development for the GAPS flight detectors. 
The GAPS prototype detectors are 2\inches-diameter and $\sim$1--1.75\,mm-thick, with a single readout strip of 1.25\inches-diameter active area. 
An example of a detector with this structure is shown in Figure~\ref{fig:testphoto}, where the active detector area is the central disk and the guard ring is the surrounding annulus. 
This active area dimension and wafer thickness were chosen in order to yield a capacitance similar to that anticipated for the flight detectors --- a key component of the ultimate noise performance (see Sec.~\ref{sec:noise}).

\begin{figure}[h]
\vspace{-0.in}
\includegraphics[width=0.6\linewidth]{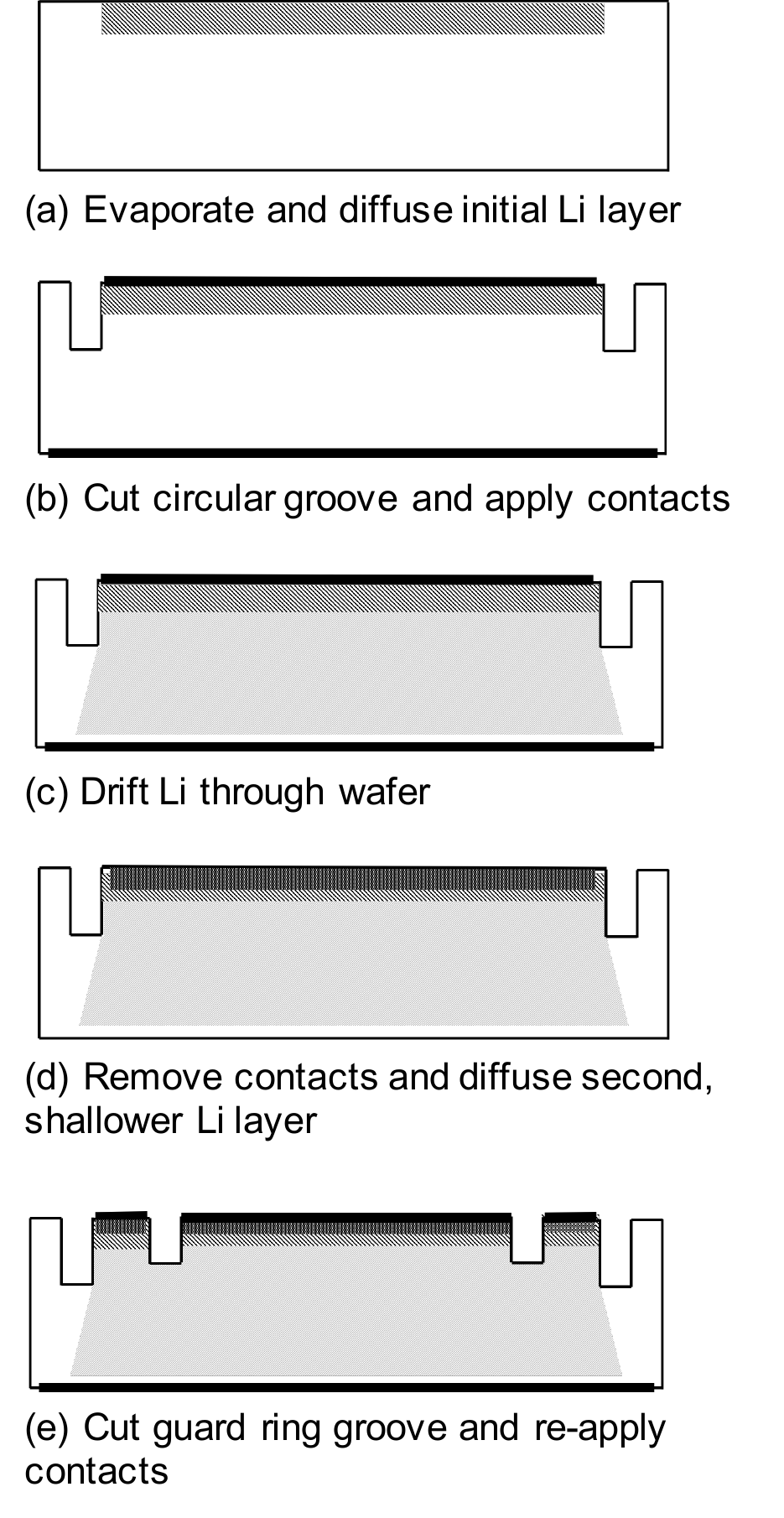}
\caption{\label{fig:fab} Overview of the prototype detector fabrication process.}
\vspace{-0.in}
\end{figure}

An overview of the prototype fabrication process is shown in Fig.~\ref{fig:fab}. 
All fabrication procedures were performed in the Columbia University Nano Initiative cleanroom, the CCNY Grove School of Engineering Cleanroom, or lab space on Columbia University campus; detector testing and assessment was performed in lab space on MIT campus and the MIT Microsystems Technology Laboratories.
The entire fabrication process for one detector can be completed in 11-12 work days, with the longest step being the drifting process, which can take up to 3-4 days depending on wafer thickness. As this time estimate includes significant ``hands-off" time for drifting, vacuum pumping, and drying of picein wax, this procedure can be efficiently parallelized for mass production.

In formulating this procedure, we have attempted to combine the best practices of the previous more than 50 years of research. 
However, as the details necessary for successful implementation were frequently lacking and were never demonstrated with our size, performance, and cost requirements, many aspects had to be re-validated in our own work. 
The procedure below resulted from the parameter study of several dozen prototype detectors.
For our most recent batch of detectors, which began as 1.75\,mm-thick wafers and underwent the exact procedure outlined below, four out of five (or a yield of $\sim$80\%) demonstrate $<10$\,nA leakage current at -40\,C, which is the threshold that allows for $\lesssim$4\,keV energy resolution (for optimized preamplifier, pulse-shaping readout electronics and pulse-shaping time, see Sec.~\ref{sec:noise}). 
Ongoing additional refinement of the production parameters for the 4\inches-diameter flight detectors, in particular the end-of-drift and etching procedures, may allow for even higher detector yield. These developments will be summarized in a future paper.
We summarize here the most important criteria we have found, as well as failure modes for each process and avenues for future investigation.

{\bf 1. Cut wafers from floating-zone (boron-doped) \ptype\ Si boules.} 
In much of the previous literature on Si(Li) fabrication, the detector geometry was limited by the size of available high-quality \ptype\ Si.
The commercial process used by the now-defunct SEMIKON Detector GmbH successfully produced large-area detectors, capable of operation at relatively high operating temperatures, but only using wafers from Topsil Semiconductor Materials ~\cite{Protic:2005}.
Specially for the GAPS experiment, SUMCO Corporation has recently developed a similar floating-zone \ptype\ substrate, the detailed characteristics of which will be described in a future paper.
A crucial outcome of our work is that wafers from both Topsil and SUMCO, with initial thickness varying from 1.25--1.7\,mm, have successfully been drifted using our procedure.
The wafers are required to have crystal orientation, bulk ingot lifetime, resistivity, and O and C impurity levels that satisfy the requirements listed in Table~\ref{tab:sispecs}. 
In addition, we require that the wafers not come from the ends of the boule, as the zone refinement process could cause the material at the ends to be inferior. 

\begin{table}[htp]
\caption{ Summary of Si substrate specifications. Detectors have been successfully fabricated using wafers from both Topsil Semiconductor Materials and SUMCO Corporation.}
\begin{center}
\begin{tabular}{cc}
\hline
Initial thickness & 1.25-1.7\,mm \\
Crystal orientation & (1-1-1) $\pm1^\circ$ \\
Bulk ingot lifetime & $>400$\,$\mu$s \\
Resistivity & 800-2000 $\Omega$-cm \\
 O impurity & $<2 \times 10^{16}$\,atoms\,cm$^{-3}$ \\
 C impurity & $<2 \times 10^{16}$\,atoms\,cm$^{-3}$ \\
 \hline
\end{tabular}
\end{center}
\label{tab:sispecs}
\end{table}%

{\bf 2. Etch clean the wafer.}
Each wafer is etched in a bath of 48\% HF, glacial acetic acid, and reagent-grade nitric acid in ratio of 4:7:11 by volume.
The reproducibility of the etch rate is enhanced by first etching a test Si wafer for approximately 5 minutes, until fuming is observed.
The detector wafer is then placed in a Teflon basket, submerged in the etchant, and gently agitated by hand for 2 minutes, taking care to lift and drain the basket every $\sim$30 seconds. 
This agitation and periodic draining allows bubbles that form on the surface of the wafer to escape, ensuring that the etchant uniformly coats the surface.   
The wafer is then quenched in a container of deionized water for 30 seconds, rinsed in running deionized water for at least 5 minutes, and dried with nitrogen.
In this process, the nitric acid forms an oxide layer on the Si, the HF removes this oxide layer, and the acetic acid adjusts the speed of the process. 
This procedure removes $\sim$7.5\,$\mu$m per minute from each exposed surface, corresponding to a total of $\sim$30\,$\mu$m in this initial etch, though the exact rate depends critically on the purity of the chemicals used. 

Contamination from copper, which is a fast diffuser in Si, could render Li drifting impossible. 
To combat this, wafers can be soaked in acetic acid for $\sim$30\,s before etching, in order to remove any traces of copper that could result from prior slicing, handling, or storage. 
However, as this step has not been found to be necessary for the wafers we process, it has not typically been applied.

{\bf 3. Evaporate Li and diffuse it into one face of the wafer.}
For this procedure, we use a custom Li evaporation system in which the wafer is held underneath and in thermal contact with a heated plate at the top of a vacuum bell jar. 
An Al mask prevents Li from evaporating onto the sides of wafer. 
The chamber is first pumped to a few $10^{-5}$\,Torr, and the heater set to 300\,C.
Once the heater plate and wafer have thermalized, the current through a tungsten boat containing $\sim10$\,g of Li, located $\sim$5.5\inches\ below the wafer, is slowly increased to 40\,A and maintained until all the Li has evaporated. 
The wafer is kept at 300\,C for total of 30 minutes, as measured from the beginning of the evaporation, and the heater plate is then rapidly cooled using a flowing ice water system.
This system was constructed primarily of off-the-shelf vacuum, heating, current source, and fluid pumping equipment, and represents a modest initial investment of $\sim$\$40k.

\begin{figure}[tp]
\vspace{-0.in}
\includegraphics[width=0.95\linewidth]{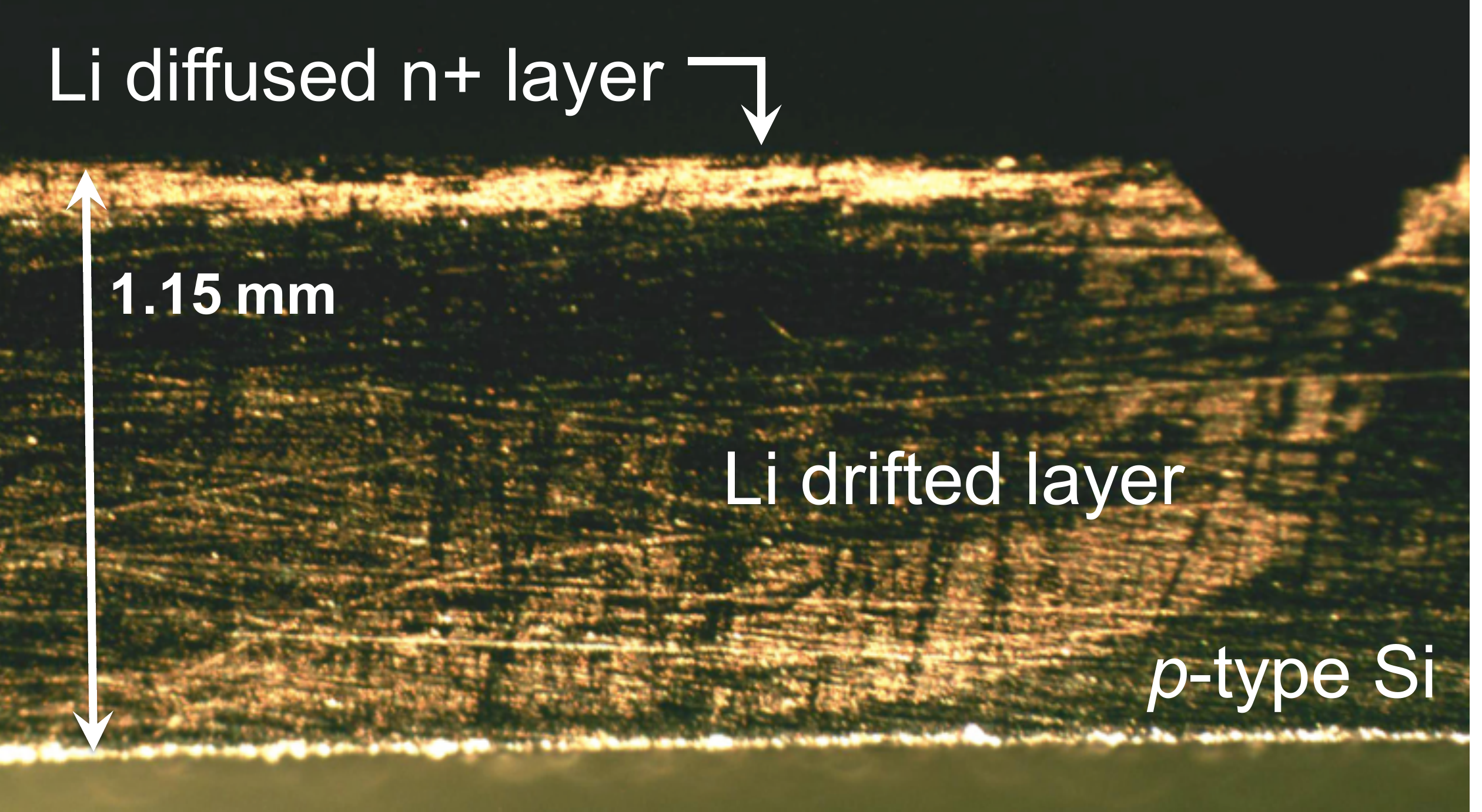}
\caption{\label{fig:copper} The cross section of a prototype detector after copper staining. The heaviest copper plating illustrates the \emph{n}+ Li diffused layer. The lighter copper plating illustrates the Li drifted region. Due to tool wear on the ultrasonic impact grinder, the groove shape is rounded and shallow, which is undesirable, as discussed in the text. The bright reflective edge on the bottom of the detector is caused by the slicing procedure.}
\vspace{-0.in}
\end{figure}

Since the exact diffused depth depends on the Li surface concentration, the B bulk concentration, the wafer temperature, and time as in Eqn.~\ref{eqn:diffusion}, 
we use copper staining, a chemical plating procedure in which copper preferentially plates where the density of Li is highest~\cite{1960BJAP...11..177I}, to visually determine the ultimate diffused Li layer depth. 
This copper staining reveals that our procedure produces a Li diffused layer $\sim$150--250\,$\mu$m deep, as shown in Fig.~\ref{fig:copper}.

Although copper staining is a crucial technique for developing and validating any Li diffusion or drifting procedure, we have found that previous references  never make clear the essential features.
In our procedure, we first prepare a saturated copper sulfate solution by adding 40\,g of copper sulfate to 100\,mL of deionized water and shaking until well combined. 
If by the following day all the copper sulfate has dissolved, we add another 5-10\,g, and repeat daily until copper sulfate crystals remain un-dissolved at the bottom of the solution. 
We then add 1\,mL of 50\% HF to this solution. 
To prepare the wafer, we cut a cross section using a diamond saw. 
Starting with 2000 diamond grit sandpaper, we sand the edge of the wafer to remove mechanical ridges that remain from the cutting procedure, taking care not to round the edges of the wafer.
This is followed by a second sanding with 2500 diamond grit sandpaper.
Sanding for at least 10 minutes total time is necessary for successful staining.

Staining is performed by pouring a thin ($\sim$1/4\inches) layer of the copper solution into a shallow Teflon dish.
We rinse the wafer briefly in deionized water, dip the desired edge into the copper solution for 12 seconds, and immediately shine a flashlight on this edge for 30 seconds. 
During this illumination, the wafer edge may still be wet, but any large droplets of solution will cause dark over-stained copper regions.
The wafer is then rinsed in running deionized water. 
In order to inspect the resultant staining, we view the edge of the wafer again with a flashlight, as the difference in copper plating is most easily visible when light reflects from the surface. 
If the desired features are not clearly visible, we sand off the copper staining and repeat the procedure.

{\bf 4. Cut a circular groove through the \emph{n}+ Li layer.}
A circular groove, either 1.7\inches\ or 1.8\inches\ in diameter, 0.5\,mm wide, and  $\sim$500~$\mu$m deep, is cut into the \nside\ of the wafer using a Raytheon Model 2-334 ultrasonic impact grinder.
This groove must be deep enough to penetrate through the \emph{n}+ layer, allowing the central region to be electrically isolated from the perimeter
and preventing Li from drifting along the side surfaces.

{\bf5. Etch the wafer and groove.} 
The wafer is etched in the HF-Nitric-Acetic mixture, following the same procedure described in Step 2, but for a shorter total etch time of 1 minute. 
This serves to clean the wafer after the ultrasonic cutting procedure, remove the heavily oxidized Li layer on the \nside, reduce any pits formed from the machining process on the groove surface, and set the proper \ntype\ surface state.

Bubbles that form during the etch process can make uniformly etching deep, narrow grooves challenging. 
For this reason, the agitation and period draining process described in Step 2 is essential when etching a detector with grooves. 
By measuring the groove width and depth using optical profilometer before and after etching, we have confirmed that this procedure yields an isotropic etch of the narrow groove surface.

{\bf 6. Apply metal contacts.}
A 1.6\inches-diameter contact is applied to the \nside\ within the boundary of the deep groove and a 1.9\inches-diameter contact is applied to the \pside. 
We have experimented with a variety of metals and application methods, typically using $\sim$40\,nm layers of either Al, Au, or Ni applied via either thermal evaporation or electon-beam vapor deposition. 
All of these contacts have allowed for successful drifting of detectors.
However, we find that using Al for both the \nside\ and \pside\ contact is preferred, as the Al allows for good adhesion with the underlying SiO$_2$ on the \nside\ and results in the clearest increase in current as the drift reaches the \pside, as described below.

{\bf 7. Drift Li from the \nside\ to the \pside\ of the wafer.}
We use a custom drifting station, monitored and controlled via a LabView interface, with the key characteristic that it maintains radial temperature variations to within $1$\,C across the diameter of our wafers in order to ensure a uniform drift speed.
This has been verified using resistance temperature detectors (RTDs) during the commissioning of the drift station equipment, and is accomplished by using a heating element that is much larger than the wafer, coupling this heater to a thermally conductive Au-coated Cu plate, and ensuring good thermal contact throughout the drift between the wafer and this heated plate. This equipment, including computer hardware and software, can be assembled for $<$\$10k.

At room temperature, we first raise the bias voltage to 250\,V and ensure that the leakage current is less than $\sim$100\,$\mu$A. 
If necessary, we leave the detector at this bias for up to several hours in order to allow localized charge concentrations resulting from non-uniformities in the Li diffusion to redistribute and the leakage current to reduce and stabilize. 
If the leakage current remains high, the contacts are masked and the grooves re-etched until the leakage current decreases below the $\sim$100\,$\mu$A threshold. 
Next, the temperature is increased gradually in 4-5 steps up to 100\,C. 
Each temperature is maintained for at least 30 minutes, or until the leakage current is no longer decreasing, before proceeding to a higher temperature.
At the final temperature, the leakage current should be less than $\sim$400\,$\mu$A for a $\sim$1.7\,mm-thick wafer.
 
\begin{figure}[tp]
\vspace{-0.in}
\includegraphics[width=0.95\linewidth]{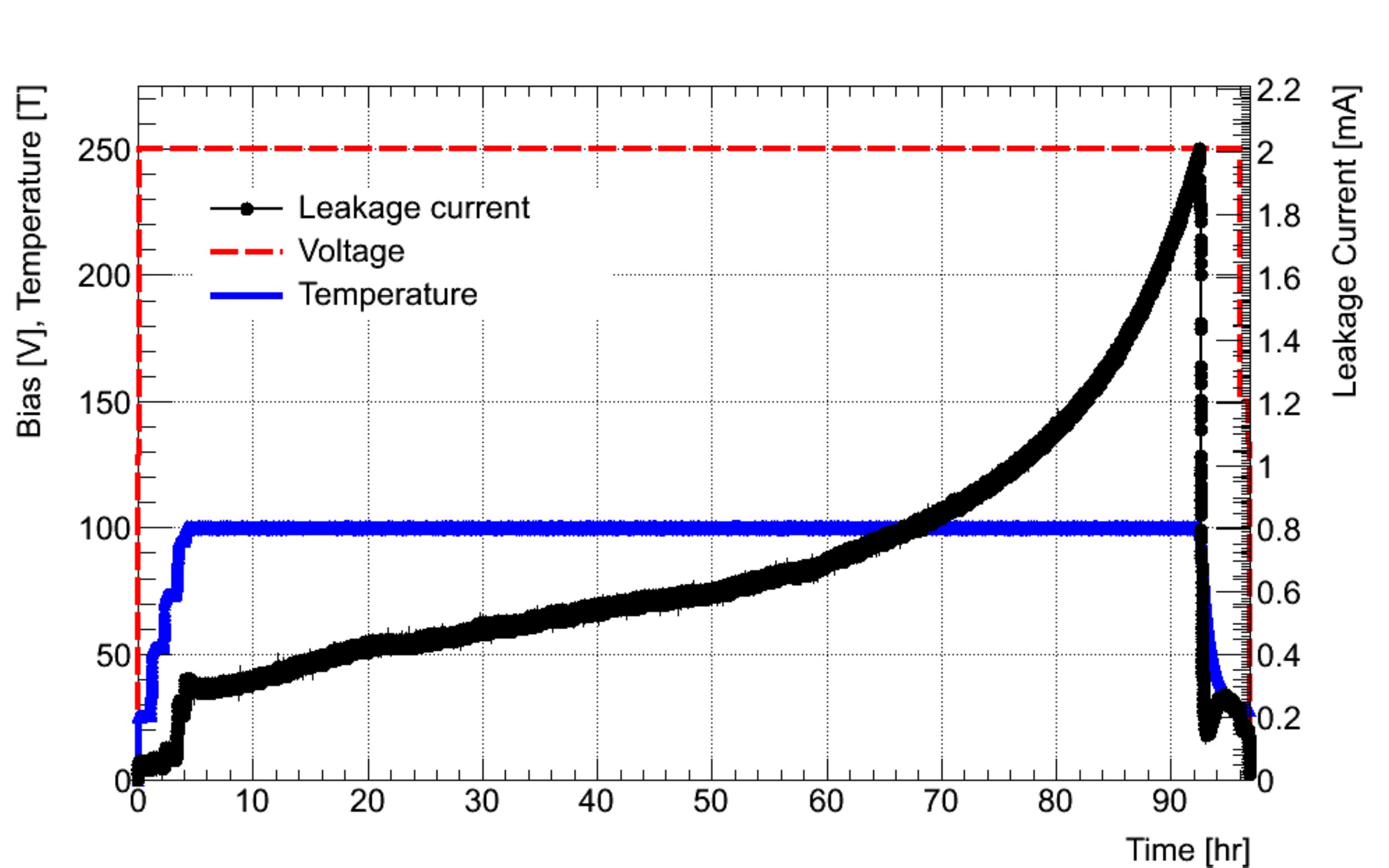}
\caption{\label{fig:drift} Bias voltage, temperature, and leakage current of TD0093 as a function of time during the drift. The sharp increase in leakage current indicates that the Li is approaching the \pside\ of the detector. When the leakage current reaches a pre-defined set-point, the LabView program automatically turns off the heater, then ramps down the bias voltage. }
\vspace{-0.in}
\end{figure}

The LabView program measures the applied voltage, temperature, and leakage current throughout the drift. 
When the leakage current increases suddenly to a predefined setpoint, 
the program turns off the heater, waits until the wafer is at approximately room temperature, and then decreases the applied bias, as shown in Fig.~\ref{fig:drift}. 
This shut-down procedure ensures that Li is not allowed to diffuse in absence of an electric field, which maintains the compensation in the bulk region. 
A  1.0\,mm drifted width requires $\sim42$~hours at 100\,C and 250\,V, while a 1.45\,mm drifted width requires 88.5 hours, with the drift beginning from the Li diffusion layer depth.
As the Li approaches the \pside, the leakage current increases more sharply for a detector with an Al Ohmic \pside\ contact than for one with a Ni or Au Schottky barrier \pside\ contact, although the Al contact must be subsequently removed and replaced with a proper Schottky barrier.

The sharp increase in current at the end of the drift indicates that the intrinsic region is approaching the \pside\ of the detector. 
Terminating the drift based only on time can be unreliable, since the drifted depth depends exponentially on temperature and it is difficult to control small temperature variations during drift. 
Our highest yield of detectors with low leakage current and good energy resolution results from ending the drift procedure at a leakage current cutoff of $\sim$2\,mA for a $\sim$1.7\,mm-thick wafer, which corresponds to a $\sim$1.75\,mm initial wafer thickness before etching.

For the prototype detectors discussed in Sec.~\ref{sec:performance}, we do not apply a cleanup drift. 
However, a cleanup drift procedure, in which the voltage bias is raised to 250\,V, the temperature to $\sim$50--70\,C, and the wafer is left for 1-2 days, has been investigated.  
As discussed in Sec.~\ref{sec:silimethod}, this cleanup drift can reduce any localized charge over-densities, particularly in the ``Li tail" region that diffuses below the concentrated \emph{n}+ region during drift. 
For some detectors, the leakage current at this bias and temperature is too high or unstable, and a cleanup drift is not possible.
For other detectors, the leakage current at room temperature and $\sim100$\,V bias is significantly (up to a factor of $10^2$) higher after the cleanup drift.
This may indicate that the cleanup drift is necessary in order to uniformly and fully over-drift the Li, but further studies are necessary.

{\bf 8. Remove contacts, mask groove, and re-evaporate a shallow Li layer onto the front surface.} 
The ultimate purpose of this second lithium evaporation is to allow for shallow segmentation of the active area and the guard ring.
Eventually, we plan to first completely remove the initial Li diffused layer and any Li ``tail" that could result from diffusion of the initial Li \emph{n}+ layer during drifting, then apply this new, thinner, Li diffused layer.
This would allow for very shallow guard ring grooves, which may further reduce surface leakage current effects.   
However, for this prototype procedure we do not remove the initial Li diffusion, as these thin wafers become too fragile if a substantial thickness of material is removed. 
Thus the procedure we describe here serves mainly to validate that we can produce very thin \emph{n}+ layers. 

We first remove the \nside\ and \pside\ contacts by lapping with 320 mesh slurry or 400 grit sand paper and de-ionized water, then polishing the surface with increasingly fine grit. 
The wafers are then ultrasonically cleaned in acetone, methanol, and de-ionized water for 3 minutes each, and etched in the HF-nitric-acetic mixture for 1 minute. 

We then use an Al mask to cover the deep groove and sides of the wafer, and evaporate a new Li layer using the same procedure as Step 3, but at a lower temperature. 
This lower temperature serves two purposes. 
First, it stably produces a uniform, thin diffusion layer. 
But also, as our system requires at least one hour to reach thermal equilibrium at the set temperature before Li evaporation, the lower temperature ensures there is no significant diffusion of the initial \emph{n}+ Li layer, growth of the Li tail, or re-distribution of Li in the drifted region, which can cause local charge non-uniformities in the bulk.
We find that a diffusion at $\sim$140~C for a total of 30 minutes produces a \emph{n}+ Li layer approximately 30\,$\mu$m thick, as verified with copper staining.

{\bf 9. Cut the guard ring groove.}
A 350~$\mu$m deep groove, 1.25\inches\ diameter and 0.5\,mm wide, is ultrasonically machined inside the outer groove. 
The region inside this groove forms the active detector area, while the region between this and the initial deeper groove forms the guard ring.
The profile of the groove shape is important to ensure that the active area and guard ring are properly electrically separated.
The profile should be as square as possible, as this will allow high electric field ``pinch points" to form, which act to increase the effective resistance across the groove~\cite{1966ITNS...13...93L}.
We measured the profile of the grooves achieved in our process using an optical profilometer and also by inspecting cross sections of a test wafer under a microscope.  
These investigations demonstrated that it is crucial to replace the tool head of the ultrasonic impact grinder periodically to avoid significant tool wear that results in more rounded groove profiles, such as shown in Fig.~\ref{fig:copper}.
We note that the diamond-slicing and sanding that is used to prepare our cross sections removes very little material, $<10$\,um, and thus does not affect the groove profile shown.

Ultimately, once we can replace the initial Li diffusion with the shallower second Li diffusion, this groove can be significantly shallower, on the order of tens of microns. Such shallow grooves can be implemented via either ultrasonic machining or ion milling techniques.

{\bf 10. Apply Ohmic contacts to the \nside.}
Electron-beam vapor deposition or thermal evaporation is used to apply $\sim$40\,nm Al Ohmic contacts to both the active area and the guard ring. We emphasize that Al is an ideal choice of metal for the \nside\, as it provides good adhesion with the underlying Li. Evaporating Au contacts onto the bare Si provides very poor adhesion, and will not survive the subsequent picein coating and removal. For the final flight detectors, though, we will over-coat the underlying \nside\ contact with Au in order to prevent oxidation.

{\bf 11. Mask \nside\ contacts and etch the grooves, sides, and \pside.}
We dissolve Apiezon brand wax in hexane or xylene until it is a smooth consistency, then paint it onto the active area and guard ring contacts, taking care to remove any wax that gets into the grooves.
The wafer is then etched in the HF-acetic-nitric mixture for 4-6 minutes.  
This sets the final surface state of the grooves and sides to be lightly \ntype.
The wax is removed using either hexane or xylene, and the wafer ultrasonically cleaned in hexane and acetone for 3 minutes each, repeated three times. 
This ultrasonic agitation is necessary to ensure the narrow grooves are uniformly cleaned. 
We have found that either hexane or xylene works for this procedure; however, xylene produces a much smoother picein application, provides more thorough removal of picein after etching, and prevents staining of the contact surface.

{\bf 12. Apply Schottky barrier contact to the \pside.}
A 1.9\inches-diameter, $\sim$100\,nm thick Ni contact is applied to the \pside\ via either thermal or electron-beam deposition. 
Many typical Si(Li) fabrication procedures use Au for Schottky barrier contact, but we have found that Au is too soft and scratches very easily. 
We have chosen Ni as the \pside\ contact, as it has a similar work function and surface barrier height as Au on Si~\cite{0022-3727-9-1-012,Sze}, but is much cheaper and more durable. 
This thick, rugged contact is necessary to prevent puncturing or scratching of any thin \pside\ barrier, which would ruin the detector as described in Sec.~\ref{sec:leakage}.
In order to improve the strength of this contact, we have attempted to overcoat Ni with Cr, but found that the interfacial stresses were too large, resulting in poor adhesion. 
As with the \nside\ contact, we plan to overcoat Ni with Au for the flight detectors, in order to prevent oxidation of the contacts.

\section{GAPS prototype Si(Li) detector performance}
\label{sec:performance}

\subsection{Experimental setup}
\label{sec:setup}

Measurements at cold temperatures are performed in a vacuum test chamber operated at $\sim3\times10^{-3}$ Torr.
This chamber provides a low-humidity environment and uses low outgassing materials, to protect the vulnerable detector surfaces. 
Detector cooling is provided via a flowing liquid nitrogen system.
The liquid nitrogen is first warmed to cold nitrogen gas by passing through copper coils outside of the chamber. 
The cold nitrogen gas then flows through 1/2\inches-diameter pipes that are mounted to an Al cold plate inside of the chamber.
Temperatures are monitored using 100\,$\Omega$ platinum RTDs encased in a ceramic casing. 
The accuracy of the RTDs was specifically calibrated for temperatures below 0\,C, and verified within $\pm1$\,C in the range -10\,C to -50\,C. 

The detector is held in a custom Al mount that routes all necessary power and signals to and from the detector while maintaining thermal contact with the Al cold plate, as shown in Fig.~\ref{fig:testphoto}. 
Negative HV bias is routed to an Al ring, on which rests the \pside\ of the detector.
Kapton tape is used to provide electrical isolation between the HV bias ring and the rest of the detector mount, while still allowing for adequate thermal conduction. 
A custom protoflight low-power low-noise charge-sensitive preamplifier is connected via kovar pin to the active area.
This preamplifier uses a 100\,M$\Omega$ feedback resistor and 0.5\,pF feedback capacitor, and requires regulated -10\,V and +6\,V DC power.
This value of feedback resistor is not optimal from a noise standpoint, and would be replaced by a 1\,G$\Omega$ value for any flight electronics.
However, for prototype studies, we use this value of feedback resistor in order to allow for diagnosis of poor-performing detectors with leakage currents above 10\,nA.
The ground of the preamplifier, common to the DC regulated power supplies, is connected to the Al cold plate via a thin copper braid, and is kept isolated from the chamber walls and any external ground sources.
The Al cold plate ground is connected to the detector guard ring via stainless steel clips, which also ensure the detector is in good thermal contact with the mount. 
During operation, a thin Al enclosure covering the entire detector and mount provides additional shielding from noise and light. 

\begin{figure}[htp]
\vspace{-0.in}
\includegraphics[width=0.8\linewidth]{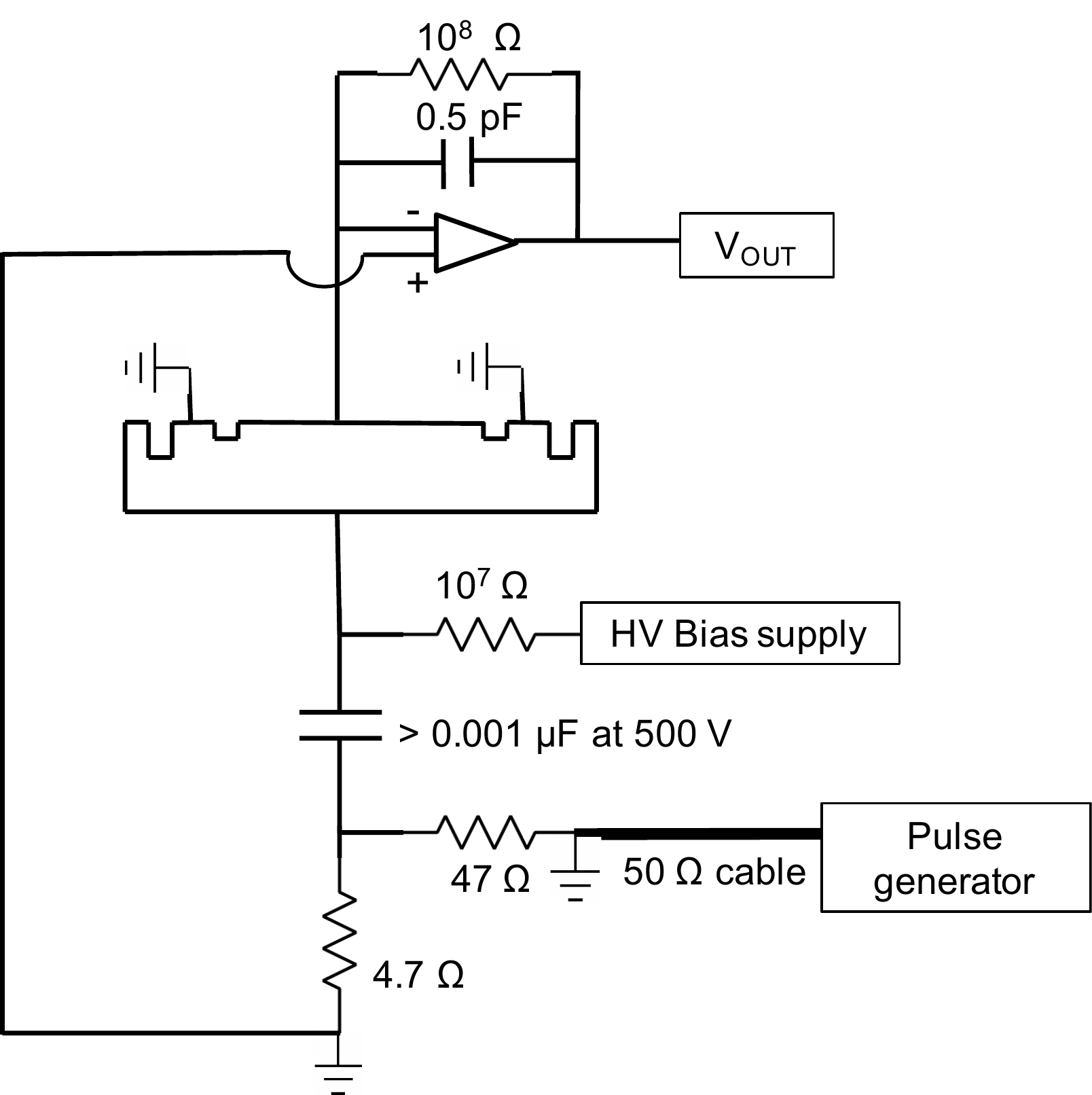}
\caption{\label{fig:testdiag} Detector power and read out for X-ray resolution testing, as described in the text. Negative HV bias is applied to the detector \pside, while the guard ring is connected to a common ground. The active area is readout from a custom charge-sensitive preamplifier. A pulse generator can be connected to provide calibration pulses. }
\vspace{-0.in}
\end{figure}

The power and readout for X-ray resolution testing is illustrated in Fig.~\ref{fig:testdiag}.
A tail pulse generator can be connected as indicated, in order to provide calibration signals to the detector. 
The 47\,$\Omega$ and 4.7\,$\Omega$ resistors shown in Fig.~\ref{fig:testdiag} provide a 11:1 voltage divider that allows a standard tail pulse generator to inject an adjustable signal over the energy range that would be  deposited by an X-ray. 
The $10^7$\,$\Omega$ resistor in the detector bias line both limits any possible very high leakage current and, in combination with the bypass capacitor, attenuates noise on the bias line. 
The preamplifier JFET is operated at a nominal 2\,mA drain current, which is less than its zero-gate-voltage drain current (I$_{DSS}$), causing the gate of the input JFET to operate slightly negative. Consequently, the preamplifier's virtual ground also operates slightly negative with respect to ground. 
When there is a path of conduction from the preamplifier input to ground, for example along the groove between the active area and the grounded guard ring, the gate of the JFET is pulled more positive, causing the preamplifier to saturate at a negative voltage.
As the detector bias is increased and the resistance increases between the guard ring and the active area, the preamplifier will reestablish the proper operating point for the virtual ground and the preamplifier output voltage will increase in a proportional positive direction.   

The preamplifier output is processed through standard NIM electronics. 
First, the signal passes through an Ortec-452 $RC-(CR)^2$ spectroscopy amplifier with a variable gain and shaping time, and measured ratio of peaking time to shaping time of 1.7. 
The signal can then be viewed on an oscilloscope or processed through a multichannel analyzer (MCA). 
For leakage current measurements at cold temperatures, the preamplifier readout is replaced with a Keithley 487 Picoammeter/Voltage source.
Testing at room temperature is performed using an Alessi REL-4500 probe station, with the Keithley 487 unit and a Boonton 7200 Capacitance meter used to measure the leakage current and capacitance.

\begin{figure}[t]
\vspace{-0.in}
\includegraphics[width=0.95\linewidth]{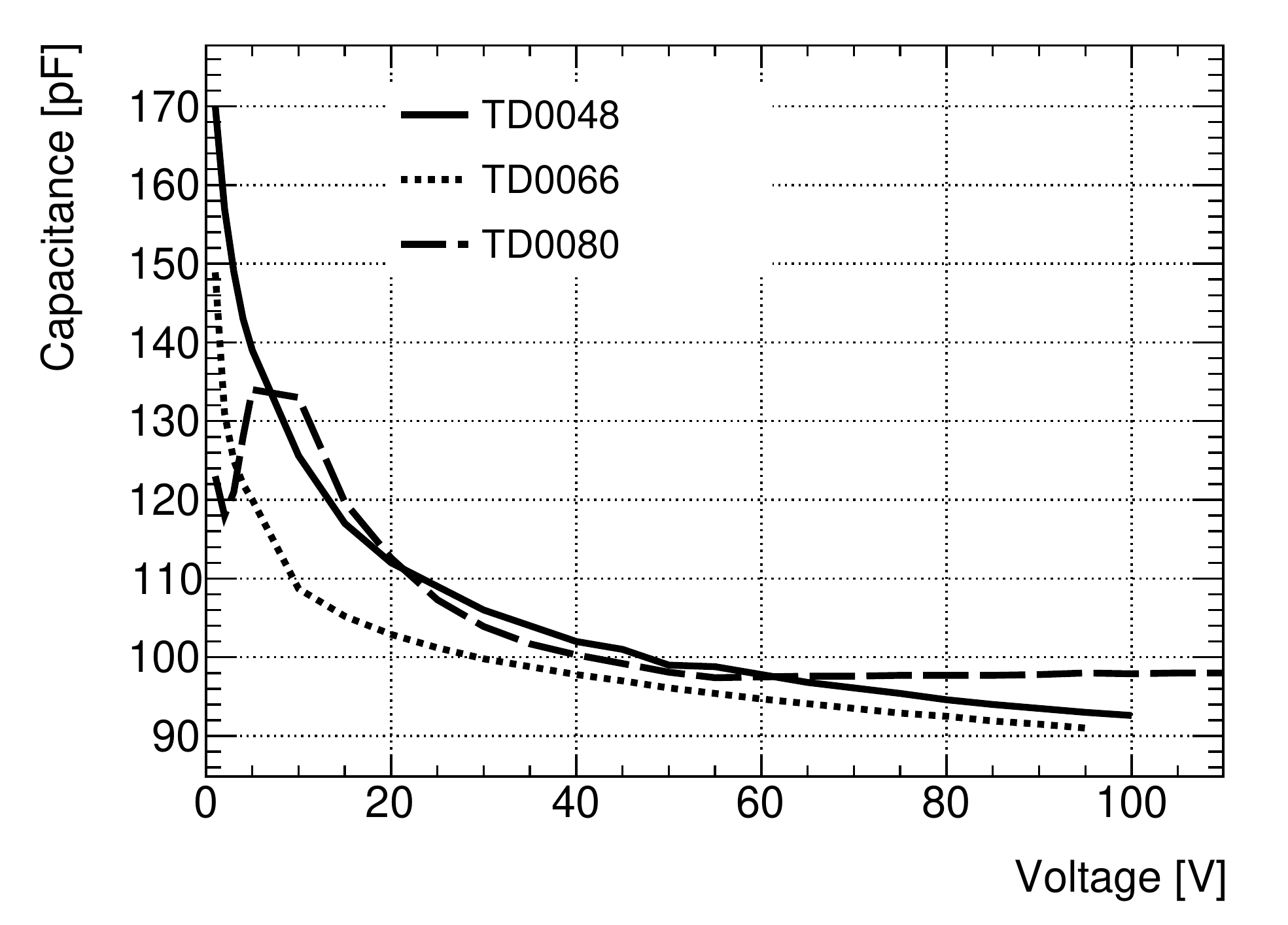}
\caption{\label{fig:cv} Capacitance as a function of bias voltage of TD0048, TD0066, and TD0080 at room temperature. The detector capacitance scales with the intrinsic region width, and is used to determine the proper operating bias. The unusual behavior of TD0080 at low bias is likely the result of the detector being unevenly and imperfectly segmented, as discussed in Sec.~\ref{sec:leakage}.}
\vspace{-0.in}
\end{figure}

\subsection{Capacitance}
\label{sec:capacitance}

Measurements of the detector capacitance, which scales with the intrinsic region width, are used to determine the operating bias at which the detector is fully depleted. 
As the reverse bias across the detector increases, the depletion region grows from the heavily-doped \emph{n}+ side toward the \pside\, and the measured capacitance decreases, as shown in Fig.~\ref{fig:cv}. 

For example, the capacitance of test detector TD0048 asymptotically approaches $\sim$90\,pF by 100\,V bias.
If the active region is modeled as a parallel plate capacitor, this corresponds to a depletion depth of $\sim$0.9\,mm. 
For this detector, which has an estimated \emph{n}+ layer depth of 150-250\,$\mu$m and a final detector thickness of 1.1\,mm, this is consistent with the detector being fully depleted. 
This calculation, however, assumes a uniform drift front, and may be inaccurate if non-uniformities in the radial temperature or electric field during drift cause some regions of the detector to be further drifted than others. 
The slight decrease in capacitance even above 100\,V bias may indicate that the depletion region is growing into the narrow Li tail region. 
In fact, as discussed below, the depletion region is thought to reach the \pside\ by $\sim$50\,V, so that any decrease in capacitance above this bias is predominantly due to either depletion of the Li tail or radially non-uniform continued growth of the depletion region.

\subsection{Leakage current}
\label{sec:leakage}

\begin{figure}[htp]
\vspace{-0.in}
\includegraphics[width=0.9\linewidth]{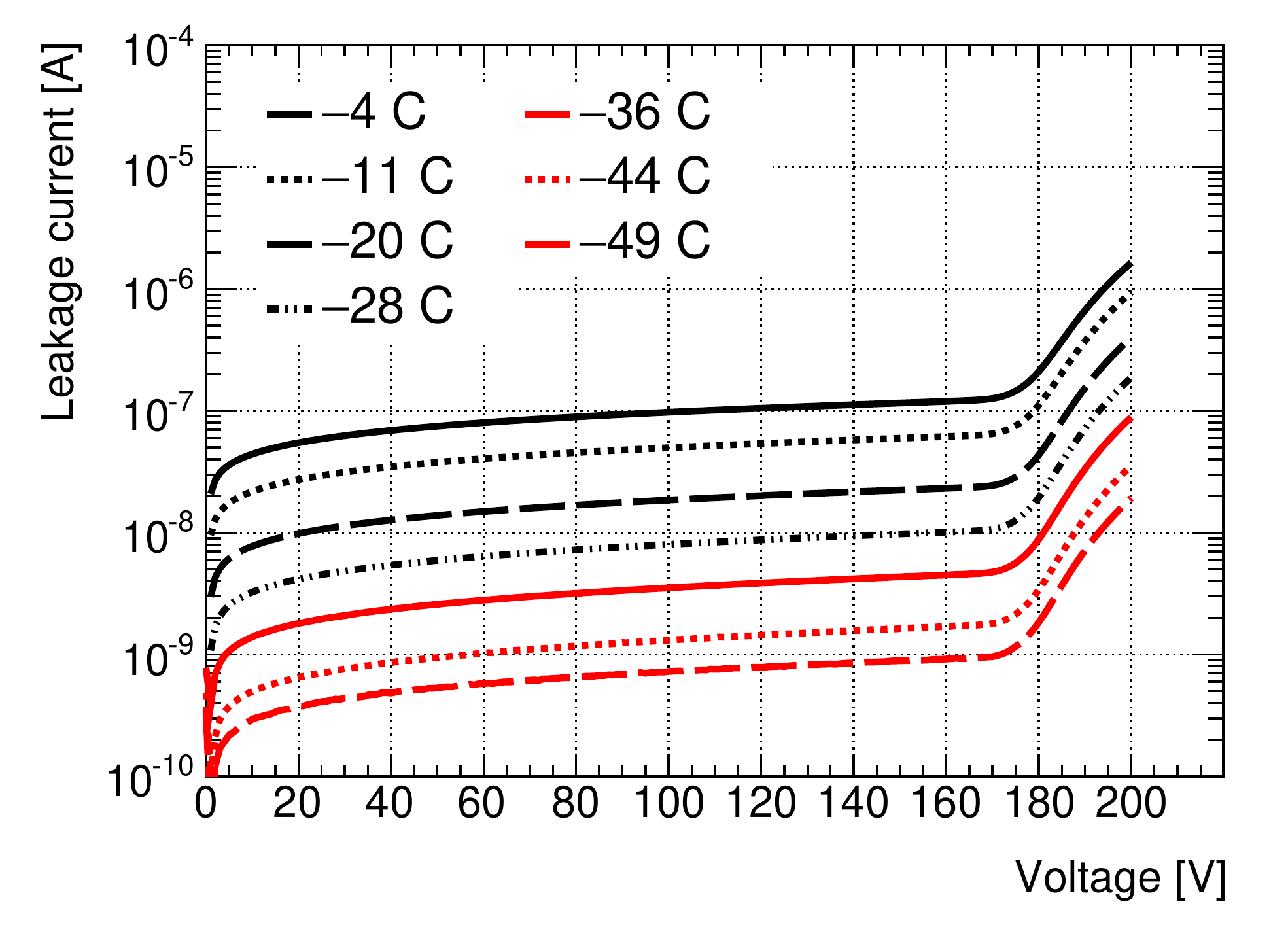}
\caption{\label{fig:iv} Leakage current as a function of bias voltage of a well-functioning detector, TD0048. The decrease in leakage current with temperature is indicative of a dominant bulk, as opposed to surface, leakage current component.}
\vspace{-0.in}
\end{figure}

\begin{figure}[htp]
\vspace{-0.in}
\includegraphics[width=0.9\linewidth]{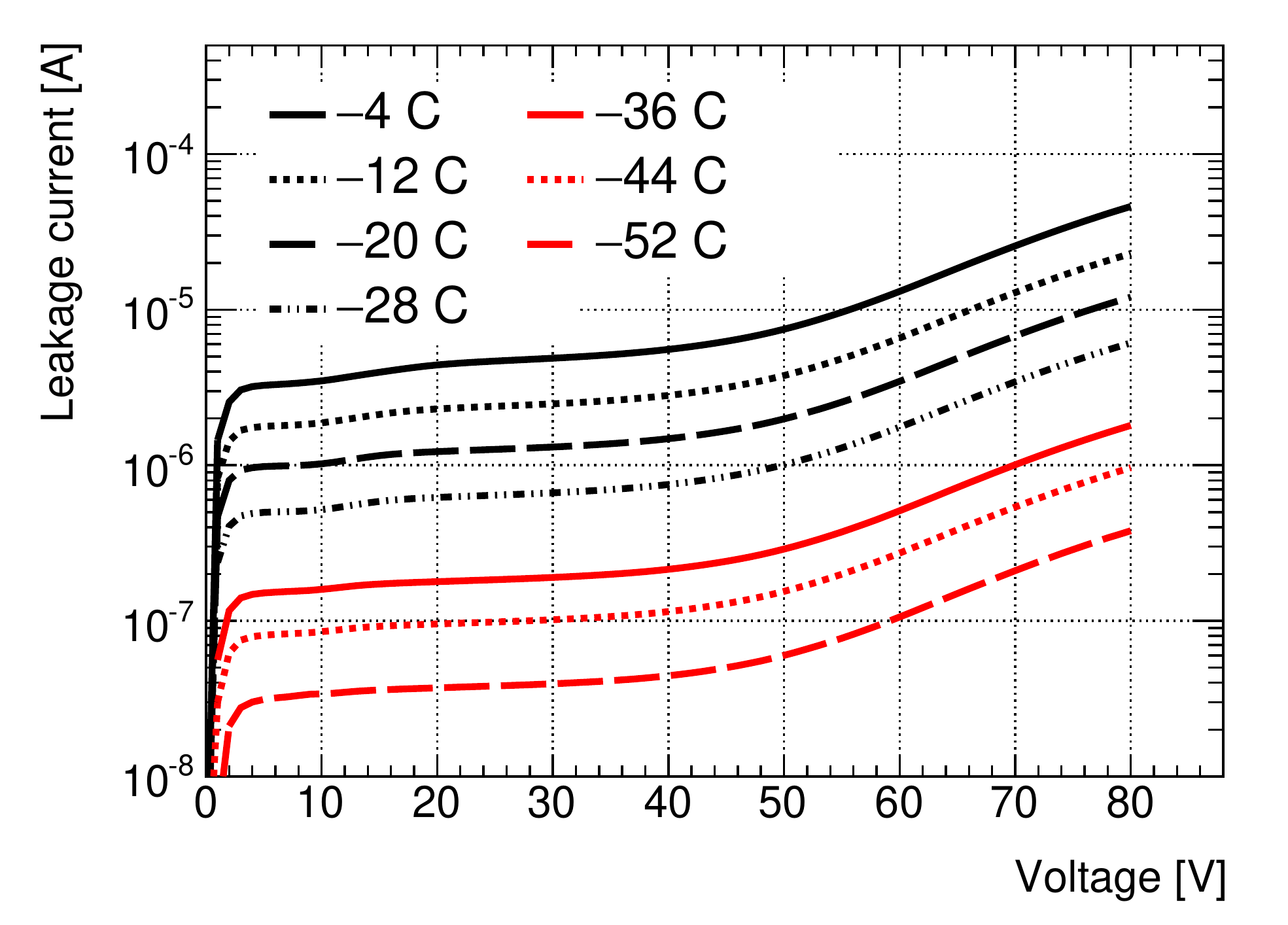}
\caption{\label{fig:badback} Leakage current as a function of bias voltage for TD0066. At low-biases, the decrease in leakage current with temperature is consistent with bulk currents, but above $\sim$50\,V bias, the leakage current increases exponentially. This is consistent with a ruptured Schottky barrier contact, where there is charge injection once the depletion region reaches the \pside\ surface.}
\vspace{-0.in}
\end{figure}

\begin{figure}[hbp]
\vspace{-0.in}
\includegraphics[width=0.9\linewidth]{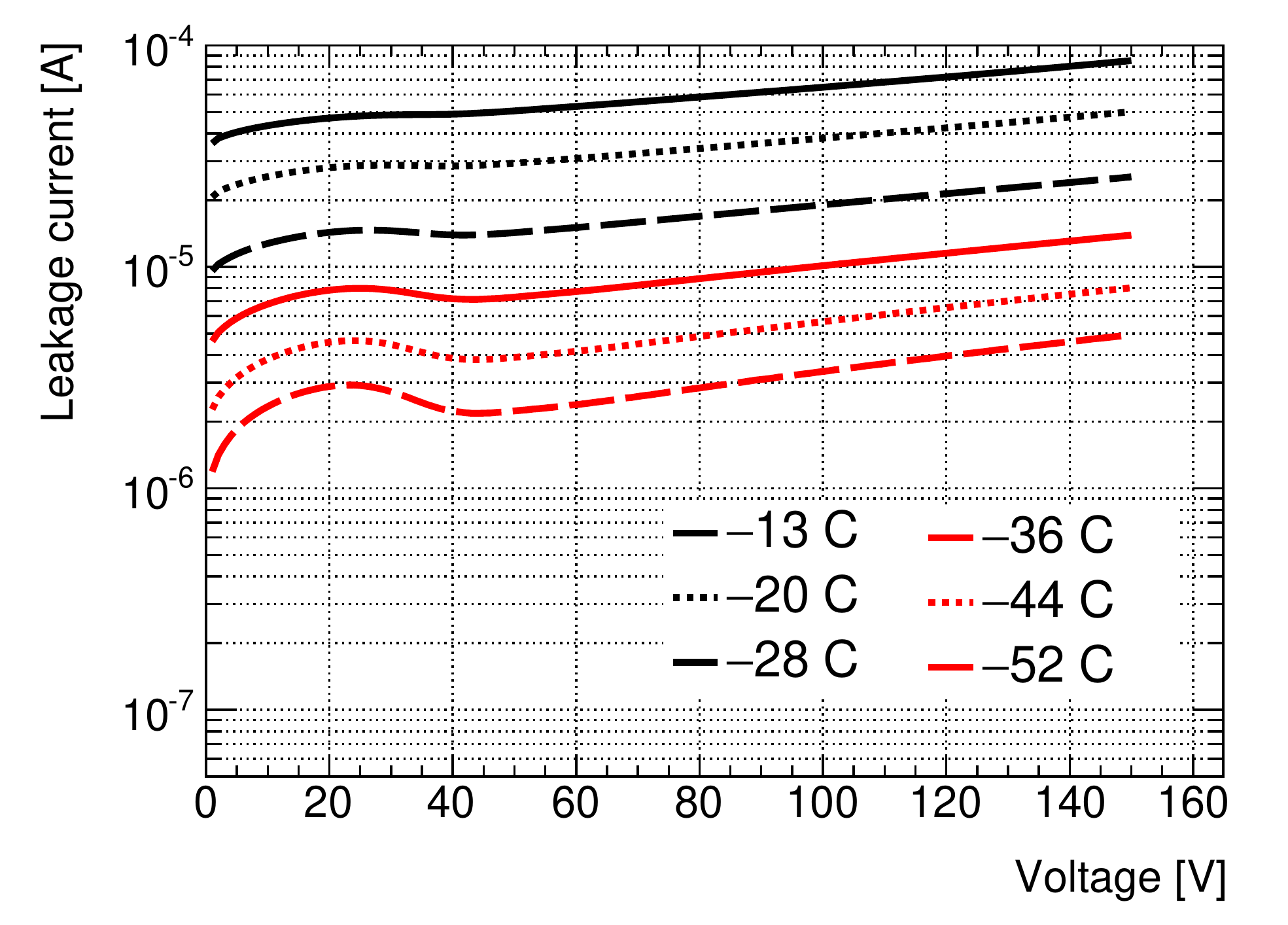}
\caption{\label{fig:badguard} Leakage current as a function of bias voltage for TD0080, a detector with a non-functioning guard ring. The leakage current is approximately diode-like, but with a magnitude several orders of magnitude higher than that of a well-functioning guard ring detector.}
\vspace{-0.in}
\end{figure}

The leakage current, or current in the absence of any true ionizing radiation signal, is a crucial parameter of the final detector energy resolution. 
The total leakage current is the sum of the bulk leakage current, which depends on the quality of the Si and the drift, and surface leakage current, which depends on the quality of the guard ring segmentation and can, if the readout strip and the guard ring do not properly separate at operating bias, be many orders of magnitude higher than the bulk current. 
For a detector dominated by bulk leakage current, the current should decrease with decreasing temperature according to $I \propto e^{-E_g/2kT}$, where $E_g$ is the energy gap in Si ($\sim1.12$\,eV), $k$ is Boltzmann's constant, and $T$ is the temperature in Kelvin~\cite{Sze}.

The leakage current as a function of reverse bias for a properly-functioning detector, TD0048, is shown in Fig.~\ref{fig:iv}. 
The leakage current decreases by approximately a factor of two for every 8\,C decrease in temperature, in accordance with the relation above. 
Note, however, that the data at the highest temperatures deviate slightly from this relation, indicating that at high temperatures the effects of poor surface states, such as caused by surface contamination, become important.
At high bias voltages, breakdown occurs in the extreme electric field and the leakage current begins increasing exponentially.

This is in contrast to the leakage currents of detectors TD0066 and TD0080, shown in Figs.~\ref{fig:badback} and \ref{fig:badguard}. 
At low biases, the TD0066 leakage current decreases with temperature as predicted for a detector dominated by bulk current, but shows an exponential increase in leakage current above $\sim$50\,V bias. 
Such behavior is indicative of charge injection as the depletion region reaches the \pside\ contact. 
This is likely caused by a ruptured \pside\ barrier, as is confirmed by visible inspection of the \pside\ surface. 
TD0080 exhibits a more extreme failure mode, with leakage currents over $10^3$ times higher than in TD0048.
Such high leakage currents indicate a non-functioning guard ring, as confirmed by subsequent inspection, which revealed that the grooves are too shallow to segment through the \emph{n}+ layer.
Thus the guard ring and the active area are not electrically separated, and surface currents dominate the leakage current.

\subsection{Energy resolution}
\label{sec:noise}

Achieving an X-ray energy resolution of $\sim4$\,keV at a temperature of -35 to -45\,C is a key requirement of the GAPS Si(Li) detectors. 
The energy resolution is dominated by shot noise from leakage current, 
thermal noise from any parallel or series resistance present in the readout chain, 
and any preamplifier noise, 
such that the equivalent noise charge ($ENC$) of a detector readout via a shaping amplifier is given by~\cite{Goulding:1966,Speiler}:
\begin{equation}
\label{eqn:noise}
ENC^2 = (2qI + \frac{4kT}{R_p})F_i \tau + 4kT(R_s + \frac{\Gamma}{g_m})F_\nu \frac{C_{total}^2}{\tau} + A_f C_{total}^2 F_{\nu f}.
\end{equation}
Here, $I$ is the leakage current and $q$ is the unit electric charge;
$R_p$ and $R_s$ are the parallel and series resistance, respectively; 
$k$ is Boltzmann's constant and $T$ is the temperature, in Kelvin; 
$g_m$ is the the transconductance of the preamplifier input stage FET, $\Gamma \approx 0.7-1$ is a constant related to the behavior of the channel in the JFET, and $A_f$ is the coefficient of the preamplifier $1/f$ noise;
$C_{total}$ is the sum of the detector capacitance, the preamplifier FET capacitance, and any stray and interelectrode capacitance;
$F_i$, $F_\nu$, and $F_{\nu f}$ are the form factors of the shaping amplifier and $\tau$ is its peaking time. 
The FWHM energy resolution is then given by:
\begin{equation}
\label{eqn:fwhm}
FWHM = 2.35 \epsilon \frac{ENC}{q},
\end{equation}
where $\epsilon$ is the energy required in Si to produce an electron-hole pair (3.6\,eV).

\begin{figure}[tp]
\vspace{-0.in}
\includegraphics[width=0.95\linewidth]{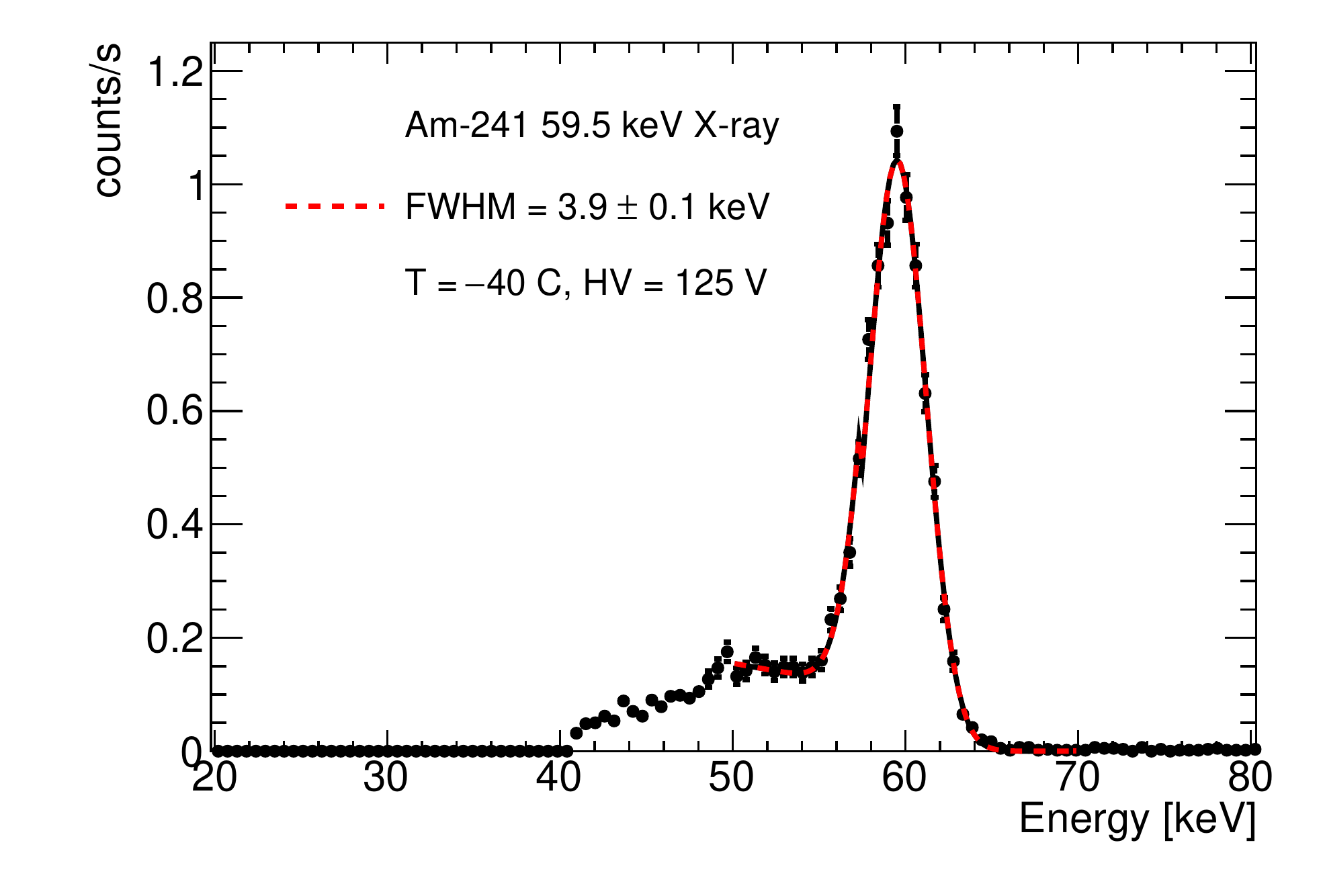}
\caption{\label{fig:spectrum} Am-241 59.5 keV line measured by TD0041 at -40\,C, 125\,V bias, and 3.4\,$\mu$s peaking time. The low-energy tail on the left side of the 59.5\,keV Am-241 peak is consistent with simulations of Compton scattering on the material inside the chamber. The peak is fit in the range 50--70\,keV to a Gaussian, with a variable-height step function describing the Compton shoulder.}
\vspace{-0.in}
\end{figure}

An example of the energy resolution for TD0041, as measured at -40\,C and 125\,V operating bias using the 59.5\,keV line of an Am-241 source, is shown in Fig.~\ref{fig:spectrum}. 
The spectrum is fit in the range 50--70\,keV to a Gaussian distribution, with mean, normalization, and standard deviation left as free parameters, and a variable-height step function describing the Compton shoulder.

\begin{figure}[htp]
\vspace{-0.in}
\includegraphics[width=0.95\linewidth]{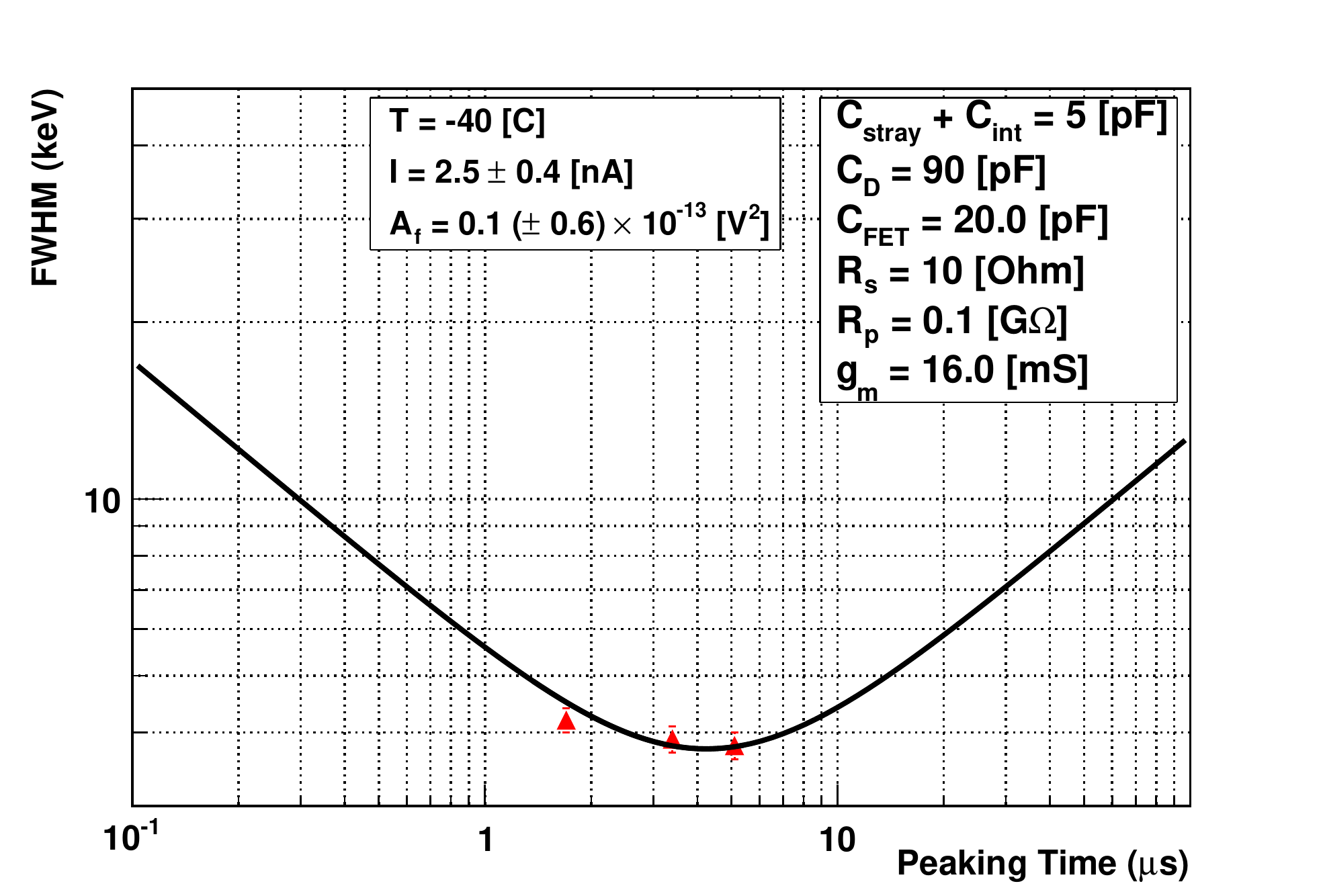}
\caption{\label{fig:noise} Energy resolution (FWHM) as a function of peaking time for TD0041 measured at -40\,C and 125\,V operating bias using the 59.5\,keV line of an Am-241 source. The black solid line shows the predicted energy resolution using the noise model described in the text, with the parameters in the inset to the right fixed to their nominal values and the best-fit leakage current and coefficient of $1/f$ noise shown in the inset to the left. The form factors for the Ortec-452 $RC-(CR)^2$ shaper are $F_i = 0.640$, $F_\nu = 0.853$, and $F_{\nu f} = 0.543$. }
\vspace{-0.in}
\end{figure}

The FWHM energy resolution as a function of peaking time for this detector is shown in Fig.~\ref{fig:noise}, with the noise model parameters shown in the inset. 
There are too few data points to constrain the many parameters of the noise model, 
so we set all parameters except the leakage current and coefficient of $1/f$ noise to nominal values.
The detector capacitance is set to the value measured at room temperature and this operating bias, and
the FET capacitance, parallel resistance, and transconductance are set to those specified by the custom preamplifier.
We use reasonable estimates for the stray capacitance and series resistance terms.
Given these choices, the data is well described by the noise model, which recovers a best-fit leakage current in agreement with those measured at the same temperature and a $1/f$ noise term that is consistent with that estimated from the nominal value of the FET and the physical input of the preamplifier. 
This shows that, for a temperature of -40\,C and peaking times of $\sim3-5$\,$\mu$s, a detector resolution of $\sim$4\,keV can be achieved. 

\section{Conclusions and future work}
\label{sec:future}

This prototype Si(Li) work has validated that detectors with sufficient performance to satisfy the unique demands of the GAPS Antarctic balloon experiment can be produced using the lithium-drifting technique for a materials cost on the order of a hundred dollars. Aside from the wet-etching facilities and thermal or electron-beam deposition equipment typically found in campus cleanrooms, the only custom equipment required are the drifting station, Li evaporator, and ultrasonic impact grinder. As described above, these require only a modest initial investment. Thanks to the low cost of materials and facilities, the required labor will be the primary cost for any large-scale production of detectors. Still, as the pGAPS detectors were produced using a proprietary commercial procedure and cost $\sim$\$18k per detector, the fabrication approach outlined above has proven that large-scale Si detector production can be accomplished within the significant cost constraints of the GAPS project.

Work now focuses on expanding this detector fabrication technique to the 4\inches-diameter, 2.5\,mm-thick, multi-strip geometry necessary for the GAPS flight detectors. 
The GAPS collaboration has formed a partnership with Shimadzu Corp., a commercial producer of Si(Li) detectors with over 30 years of experience, who will use their facilities and skilled technical staff to produce the $\sim$1000 flight detectors.
Like other commercial producers in business today, Shimadzu specializes in much smaller (1\,cm-diameter) detectors that
operate at much lower temperatures ($<$-150\,C).
Because of this, we are now working to merge the GAPS prototype detector techniques with the Shimadzu detector technology and to optimize for the larger geometry. 
In particular, this optimization focuses on improving detector segmentation and strip yield, reducing detector dead area, and developing long-term surface passivation~\cite{1994NIMPA.353...89J} methods. 
Full-scale fabrication of GAPS flight detectors scheduled for 2018-2020, with the initial GAPS Antarctic balloon flight scheduled for 2020-2021. 

\section*{Acknowledgements}
This work was supported in part by NASA APRA grant NNX17AB44G, and by MEXT/JSPS KAKENHI grants JP26707015 and JP17H01136.
K. Perez receives support from the Heising-Simons Foundation and RCSA Cottrell College Science Award ID \#23194.
This material is based upon work supported by the National Science Foundation under Award No. 1202958 and through the Graduate Research Fellowship under Grant No. 1122374. 
We thank SUMCO Corporation for their collaboration to develop Si material specially for the GAPS experiment. 

\section*{References}
\bibliography{Perez_SiLiGAPS}

\end{document}